\documentclass[runningheads]{llncs}
\usepackage{graphicx}
\usepackage{wrapfig}
\usepackage{subcaption}
\usepackage{float}
\usepackage{amsmath}
\usepackage{mathtools}
\usepackage[english]{babel}
\usepackage{cite}
\usepackage{textcmds}
\usepackage[utf8]{inputenc}
\usepackage{lineno}
\usepackage[official]{eurosym}

\begin{document}
\title{Systems Mining with {\normalfont \textsc{Heraklit}}: The Next Step}
%
%

\author{Peter Fettke\inst{1,2}\orcidID{0000-0002-0624-4431} \and
Wolfgang Reisig\inst{3}\orcidID{0000-0002-7026-2810}}
\authorrunning{P. Fettke, W. Reisig}
%

\institute{German Research Center for Artificial Intelligence (DFKI), Saarbr\"ucken, Germany \\
\email{peter.fettke@dfki.de}\\ \and
Saarland University, Saarbr\"ucken, Germany \\ \and
Humboldt-Universität zu Berlin, Berlin, Germany \\  
\email{reisig@informatik.hu-berlin.de}}

\maketitle              
\begin{abstract}
We suggest \textit{systems mining} as the next step after process mining.  Systems mining starts with a more careful investigation of runs, and constructs a detailed model of behavior, more subtle than classical process mining. The resulting model is enriched with information about data. From this model, a system model can be deduced in a systematic way. 

\end{abstract}

\section{Introduction}

Classical process mining methods as established in theory and practice start out with \textit{event logs}, generated by \textit{processes} during their dynamic progression \cite{aalst2016mining,gartner2019market_guide}. Process mining is designed first of all to discover processes by extracting knowledge from event logs. Each event in an event log is conceived as an activity that has been performed in the process at the point in time given in the event log, and is related to a particular case. Typically, the events of a case are totally or weakly ordered and can be seen as an execution or run of the process. 

\begin{figure}[!tb]
\centering
\includegraphics[scale=.25]{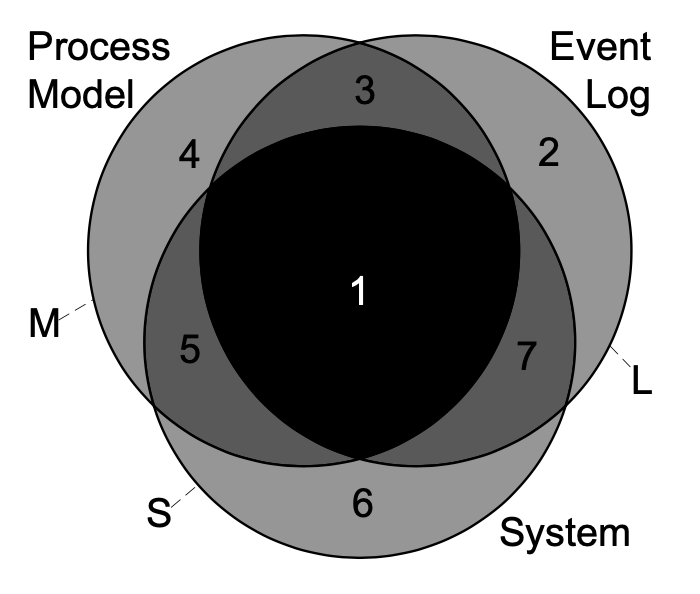}
\includegraphics[scale=.15]{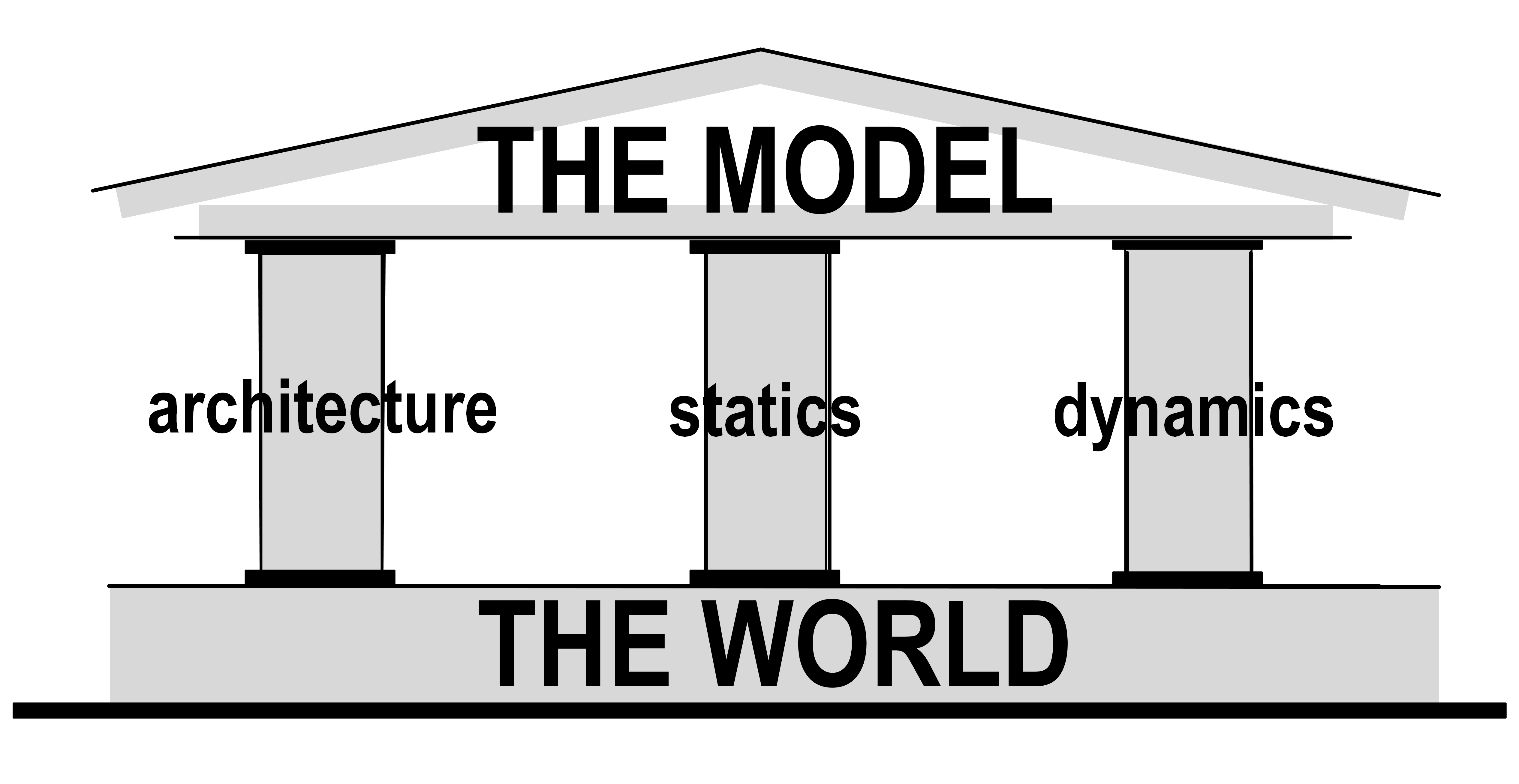}
\caption{classical process mining (left, source: \cite{buijs2014quality_dimensions}) and systems mining.}
\label{fig:heraklit_temple}
\end{figure}

The left side of Fig.~\ref{fig:heraklit_temple} depicts the standard formal approach for understanding an event log, a processes model, and a system, namely, behavior can be understood as three different sets of symbol sequences \cite{buijs2014quality_dimensions}. In this paper, we propose to follow a different route: there is no reason to assume that the events of a run are totally or weakly ordered. Of course, a clock outside the run may timestamp a run’s events. This induces an order; however, this order is irrelevant for a proper understanding of a run. To the contrary, it spoils the \textit{causal} order of events, which orders two events $a$ and $b$ by $a < b$ if and only if $a$ is a prerequisite for $b$. Of course, $a < b$ implies each potential clock to timestamp $a$ before $b$.  But $a$ timestamped before $b$ only implies that $b$ is not a prerequisite for $a$. Or, in one sentence: causality matters!

Additionally, systems to be mined are typically not monolithic, amorphous or unstructured, but can best be described and understood as the composition of different sub-systems. Hence, an understanding of the different modules of a system is necessary while understanding the behavior of the system. Again, in one sentence: composition matters!

Last, but not least, a system processes data. Data processing is not only needed for the correct execution of processes, but also for the symbolic representation of important objects, e.g. invoices, customers, agreements, orders, products, and many more objects of interest. These objects need to be understood and represented adequately while mining a system. Again, in one sentence: objects matter!

In this contribution, by means of an example, we show how to mine not only process models, but entire system models. This includes the integrated modeling of architecture, statics, and dynamics of the world we live in (Fig.~\ref{fig:heraklit_temple}, right). To this end, we combine the well-known techniques of Petri nets and abstract data types with the recent \textit{composition calculus}.

We motivate and exemplify a different notion of \textit{runs}, by means of a case study from the area of retail sale. Supported by some static aspects of a system, such runs can be deduced from the system’s event logs. Note, that our paper is purely conceptual. We do not provide an algorithm nor a software tool for systems mining. Instead, our main contribution is the elaboration of the new idea of systems mining based on the formal framework of \textsc{Heraklit} \cite{fettke2021modelling,fettke2021handbook}.

This paper starts with the presentation of the main idea of modules and runs while unfolding a running case study (Sec.~2). Sec.~3 describes systems nets and Sec.~4 presents the main idea for mining a system module. Related work is discussed in Sec.~5, Sec.~6 presents some conclusions.

\section{Modules and Their Composition}
Before presenting the (not too heavy) formal framework, we discuss a motivating example that later will be extended to a full case study.

\subsection{Example: Occurrence Modules of a Retail Business}

\begin{figure}[!tb]
\centering
{\includegraphics[trim={0cm 0cm 0cm 0cm},clip,scale=.25]{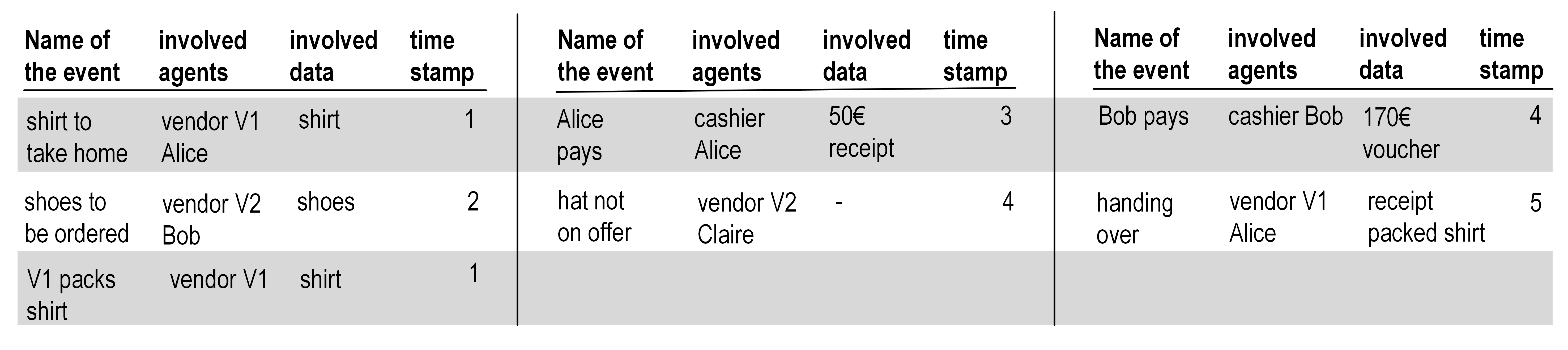}}
\caption{event log with seven events.}
\label{fig:event_log_neu}
\end{figure}

We start with a small log, recording observations from the field, namely seven events from a retail shop, as Fig.~\ref{fig:event_log_neu} shows. Each event has a unique name, a set of involved agents, a set of data, and a timestamp. Static inspection of the system and the events of the log identifies \textit{six agents}: Two vendors $V1$ and $V2$, a cashier, and three clients, Alice, Bob, and Claire. All events of the log, up to \textit{$V1$ packs shirt}, include two agents. For example, the \textit{shirt to take home} event includes the vendor $V1$ and the client Alice, jointly selecting a shirt for Alice. The \textit{shoes to be ordered} event includes the vendor $V2$ and the client Bob, jointly agreeing on shoes, to be ordered from wholesale. The other events are intuitively obvious.

\begin{figure}[!tb]
\centering
{\includegraphics[trim={0cm 0cm 0cm 0cm},clip,scale=.25]{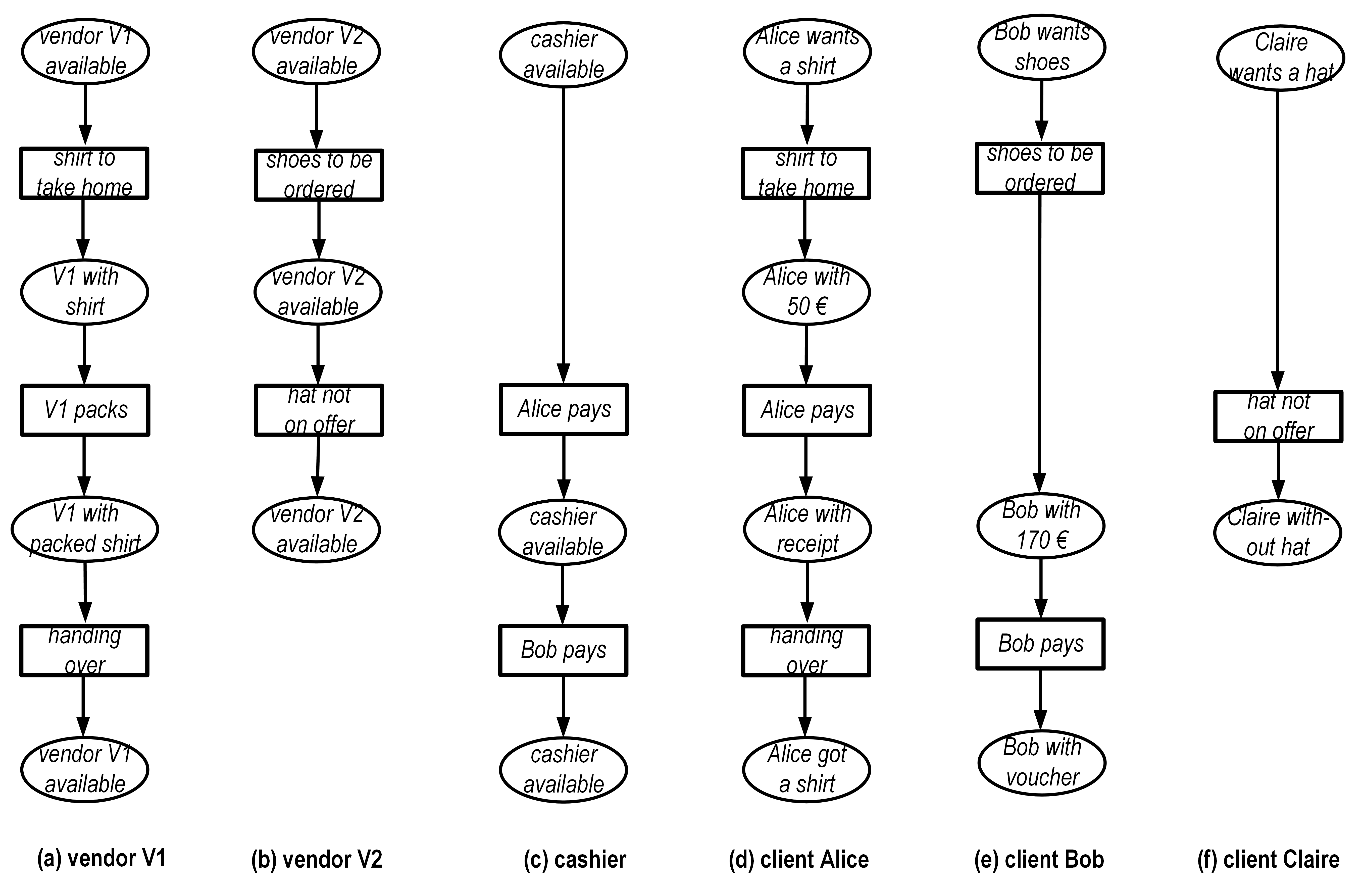}}
\caption{agents and their behavior, elicited from the event log.}
\label{fig:six_behaviours}
\end{figure}

From the perspective of agents it is intuitively obvious that for a given event log, an agent is involved in a \textit{sequence} of events, describing one of the potential \textit{behaviors} of the agent. For example, the event log in Fig.~\ref{fig:event_log_neu} implies the vendor $V1$ be involved in three events: \textit{shirt to take home}, \textit{V1 packs shirt}, and \textit{handing over}. This behavior can automatically be deduced from the log. An event updates the local state prior to its occurrence, and produces a local state as a result of its occurrence. Technically, we represent this as a Petri net, as in Fig.~\ref{fig:six_behaviours} (a). Each place (circle) denotes a local state; each transition (rectangle) denotes a step. An agent’s behavior deduced from a log is very simple in structure; it can be thought of as a classical sequence of states and steps.

\begin{figure}[!tb]
\centering
{\includegraphics[trim={0cm 0cm 0cm 0cm},clip,scale=.25]{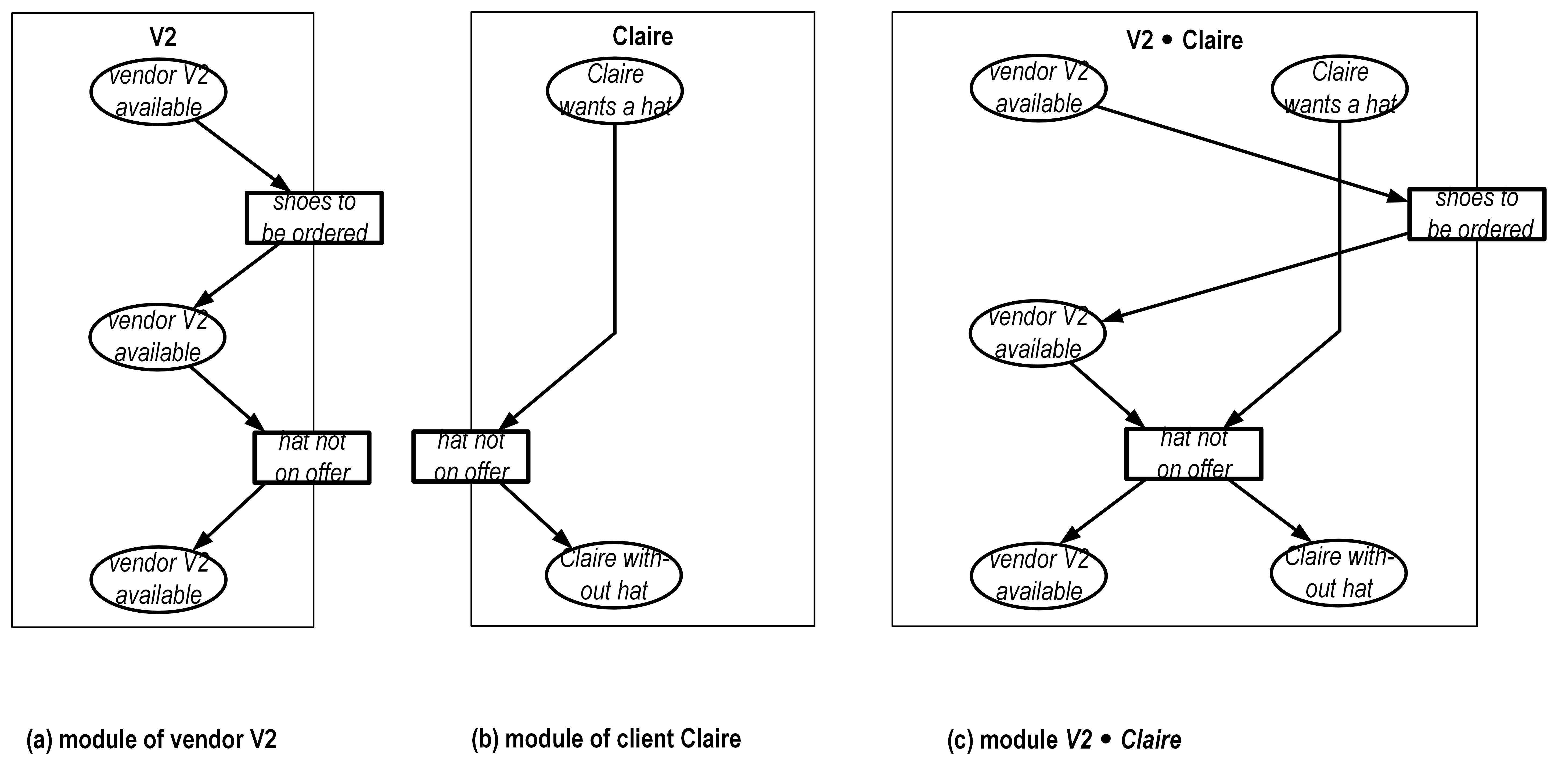}}
\caption{two modules and their composition.}
\label{fig:V2-Claire}
\end{figure}

Fig.~\ref{fig:six_behaviours} shows the behaviors of all six agents, deduced from the log in Fig.~\ref{fig:event_log_neu}. Obviously, they are tightly interrelated, and this interrelation is now to be constructed explicitly. To do this, each behavior is embedded into a \textit{module} in which each transition is either inside the module, or in an \textit{interface} of the module. Each module has a \textit{left} and a \textit{right} interface. Graphically, a module is enclosed in a rectangle with the left and right interface elements on the left and right margin, respectively. The left and right interface of a module $A$ is designated $^\ast \! A$ and $A^\ast$, respectively. This way, Fig.~\ref{fig:V2-Claire} (a) shows the module $V2$ of the vendor $V2$. The two transitions of the module are both located on the right interface, $V2^\ast$. The \textit{Claire} module in Fig.~\ref{fig:V2-Claire} (b) places the transition of Claire's module on the left interface, $^\ast \! \textit{Claire}$.

The two modules are now composed into a new module, $V2 \bullet \textit{Claire}$, shown in Fig.~\ref{fig:V2-Claire} (c). To compose $V2$ and \textit{Claire}, we merge the transition with label \textit{hat not on offer} of $V2^\ast$ with the equally labeled transition of $^\ast \! \textit{Claire}$. The resulting transition goes inside $V2 \bullet \textit{Claire}$. The transition with label \textit{shoes to be ordered} of $V2^\ast$ goes to $(V2 \bullet \textit{Claire})^\ast$.  

This example shows the general principle of the composition of two modules $A$ and $B$: Equally labeled elements of $A^\ast$ and $^\ast \! B$ are merged and go into the interior of $A \bullet B$. The other elements from $A^\ast$ and $^\ast \! B$ go to $(A \bullet B)^\ast$ and $^\ast \! (A \bullet B)$, respectively. This kind of composition motivates the distinction of right and left interfaces: The running example exhibits an intuitive dichotomy between shop modules (vendors and cashiers), and client modules. Shop modules interact with client modules, so the interface elements of shop modules and of client modules complement each other. 

\begin{figure}[!tb]
\centering
\begin{minipage}{.6\textwidth}
{\includegraphics[trim={0cm 0cm 0cm 0cm},clip,scale=.20]{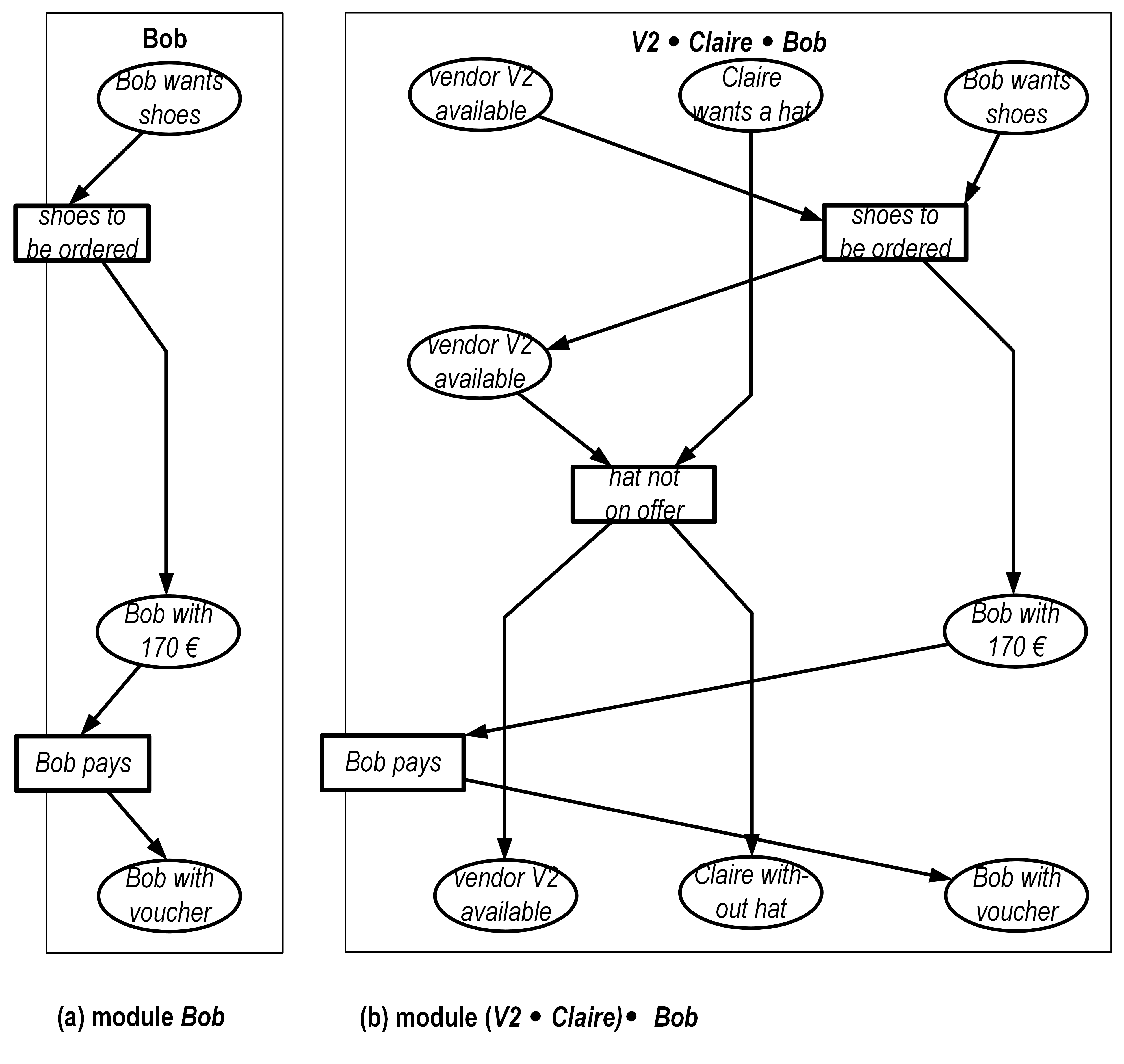}}
\caption{composing another module.}
\label{fig:V2-Claire-Bob}
\end{minipage}%
\begin{minipage}{.3\textwidth}
{\includegraphics[trim={0cm 0cm 0cm 0cm},clip,scale=.20]{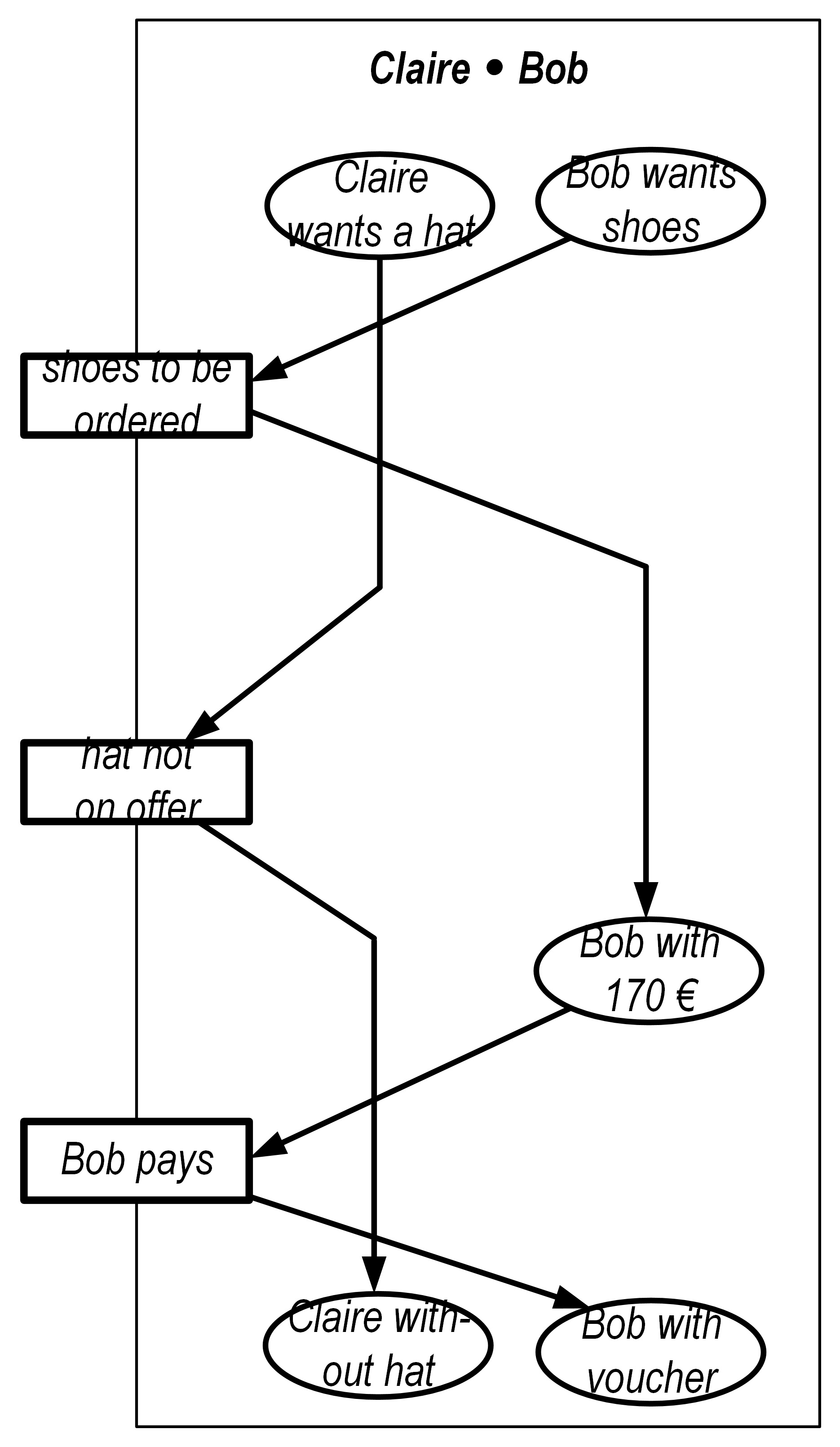}}
\caption{Claire $\bullet$ Bob.}
\label{fig:Claire-Bob}
\end{minipage}%
\end{figure}

To continue, Fig.~\ref{fig:V2-Claire-Bob} (a) shows the module \textit{Bob} of the client Bob from Fig.~\ref{fig:six_behaviours} (e). As with the \textit{Claire} module, its transitions lie in its left interface. We now compose $V2 \bullet \textit{Claire}$ with \textit{Bob} and obtain the module $(V2 \bullet \textit{Claire}) \bullet Bob$ in Fig.~\ref{fig:V2-Claire-Bob}(b).  Alternatively, we could have formed the module $\textit{Claire} \bullet \textit{Bob}$ first (Fig.~\ref{fig:Claire-Bob}) and then module $V2 \bullet (\textit{Claire} \bullet \textit{Bob})$. It is easy to see that the modules $(V2 \bullet \textit{Claire}) \bullet \textit{Bob}$ and $V2 \bullet (\textit{Claire} \bullet \textit{Bob})$ are identical. We will see that in general, the composition operator $\bullet$ is \textit{associative}.

\subsection{The Formal Framework of Modules}

As usual, we represent a Petri net as a triple $(P,T; F)$. We employ the usual graphical representation with boxes, circles, and arrows. In this section, we recall a special case of the composition calculus, and particularly occurrence modules and their composition. The general case can be found in \cite{fettke2021handbook}.
 
An \textit{interface over} a set $\Lambda$ \textit{of labels} is a finite set $R$, with each element of $R$ carrying a label of $\Lambda$. We refrain from the general case of two or more equally labeled interface elements here.

For two interfaces $R$ and $S$, equally labeled elements $r \in R$ and $s \in S$, are a \textit{harmonic pair} of $R$ and $S$. A harmonic pair is labeled by the label of $r$ and $s$. The element $s$ is a \textit{harmonic partner} of $r$ in $S$, and $r$ is a \textit{harmonic partner} of $s$ in $R$.

A \textit{module} is a Petri net $N = (P, T; F)$ together with two interfaces $^\ast \! N$ and $N^\ast \subseteq P \cup T$, denoted as the \textit{left} and the \textit{right interface} of $N$. Nodes not in an interface belong to the \textit{interior of} $N$.

In graphical representations, the interior of $N$ is surrounded by a box, with the elements of the left and the right interface on its left and the right margin, respectively, e.g. Fig.~\ref{fig:V2-Claire}.

We are now prepared for the fundamental definition of \textit{composing} two modules: 

Let $A$ and $B$ be two modules. For each node $x$ of $A$ or of $B$, let $x' = \{x, y\}$ if $\{x, y\}$ is a harmonic pair of $A^\ast$ and $^\ast \! B$; let $x' = x$ if no harmonic pair of $A^\ast$ and $^\ast \! B$ contains $x$. Then the module $A \bullet B$ is defined as follows (each element retains its label):

\begin{enumerate}

\item The nodes of $A \bullet B$ are all $x'$ such that $x$ is a node of $A$ or of $B$.

\item The edges of $A \bullet B$ are all $(x', z')$, such that $(x, z)$ is an edge of $A$ or of $B$.

\item The left interface $^\ast \! (A \bullet B)$:

\begin{enumerate}
\item $^\ast \! A \subseteq {^\ast (A \bullet B)}$;
\item For $x \in {^\ast B}$ holds: $x \in {^\ast (A \bullet B)}$, if $x$ has no harmonic partner in $A^\ast$.
 \end{enumerate}

\item The right interface $(A \bullet B)^\ast$:

\begin{enumerate}
\item $B^\ast  \subseteq (A \bullet B)^\ast$;
\item For $x \in A^\ast$ holds: $x \in (A \bullet B)^\ast$, if $x$ has no harmonic partner in $^\ast \! B$.
\end{enumerate}

\end{enumerate}

Figs.~\ref{fig:V2-Claire}, \ref{fig:V2-Claire-Bob} etc. show compositions of modules. Notice that, according to this definition, $^\ast \! (A \bullet B)$ or $(A \bullet B)^\ast$ may acquire different elements with equal labels. However, this never happens in this paper’s examples; further details can be found in \cite{fettke2021handbook}.

A fundamental property of composition is \textit{associativity}, decisive for the usability of modules and their composition. In fact, for any three modules $A$, $B$ and $C$ holds:
\begin{equation}
    (A \bullet B) \bullet C  =  A \bullet (B \bullet C).
\end{equation}
As a consequence, it makes sense to just write $A \bullet B \bullet C$. This property is a special case of a more general notion of modules and their associative composition, as discussed in \cite{reisig2019associative}.
 
Furthermore, there are clear criteria for the case of commutativity: for modules A and B holds
\begin{equation}
    A \bullet B  =  B \bullet A
\end{equation}
if and only if no label occurs in $^\ast \! A \cup A^\ast$ as well as in $^\ast \! B \cup B^\ast$.

The nets in the examples of Sec. 2.1 all exhibit a particular structure: The arcs form no cycles, and each place has at most one ingoing and one outgoing arc:

A net $N = (P, T; F)$ is an \textit{occurrence net} if and only if:
\begin{enumerate}
\item The transitive closure of $F$, usually written as $F^+$, is a strict partial order, viz. irreflexive and transitive, on $P \cup T$. We denote this relation as $<_N$; 
\item for each $p \in P$ there exists at most one arc shaped $(t, p)$ and at most one arc shaped $(p, t)$.
\end{enumerate}
A module is an \textit{occurrence module} if and only if the underlying net is an occurrence net.

\begin{figure}[!tb]
\centering
{\includegraphics[trim={0cm 0cm 0cm 0cm},clip,scale=.20]{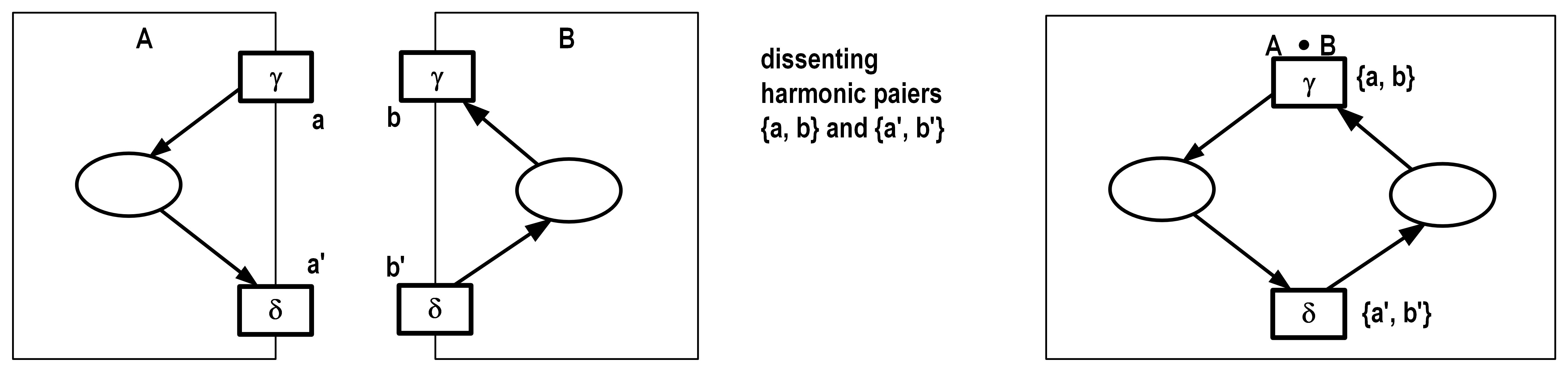}}
\caption{dissenting pairs.}
\label{fig:dissenting_pairs}
\end{figure}

For two occurrence modules $A$ and $B$, the composed module $A \bullet B$ is in general not an occurrence module again. Fig.~\ref{fig:dissenting_pairs} shows an example. This example shows that the interior of $A$ and $B$ matters for this problem. Nevertheless, it can be reduced to a problem of the induced order of interface elements: 

With $a$, $a' \in A$ and $b$, $b' \in B$, let $\{a, b\}$ and $\{a', b'\}$ be harmonic pairs of $A^\ast$ and $^\ast \! B$. They \textit{dissent} if and only if either $a <_A a'$ and $b' <_B b$, or $a' <_A a$ and $b <_B b'$.

Then, for two occurrence modules $A$ and $B$ it holds: $A \bullet B$ is an occurrence module if and only if $A^\ast$ and $^\ast \! B$ have no dissenting harmonic pairs. All compositions of occurrence modules in this paper yields an occurrence module again.

\subsection{Completing the Example}
\begin{figure}[!tb]
\centering
{\includegraphics[trim={0cm 0cm 0cm 0cm},clip,scale=.20]{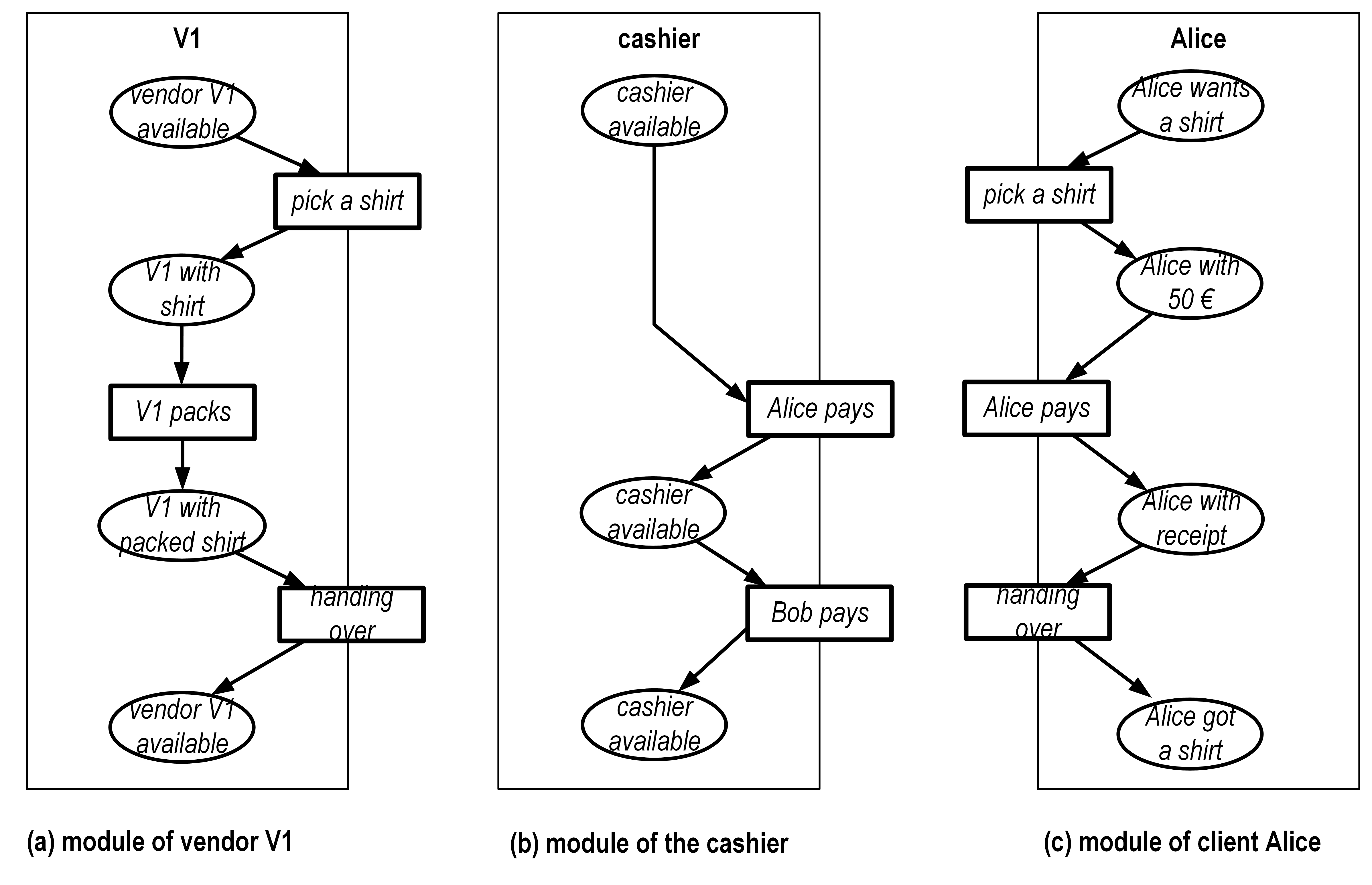}}
\caption{three further modules.}
\label{fig:V1-cashier-Alice}
\end{figure}

\begin{figure}[!tb]
\centering
{\includegraphics[trim={0cm 0cm 0cm 0cm},clip,scale=.20]{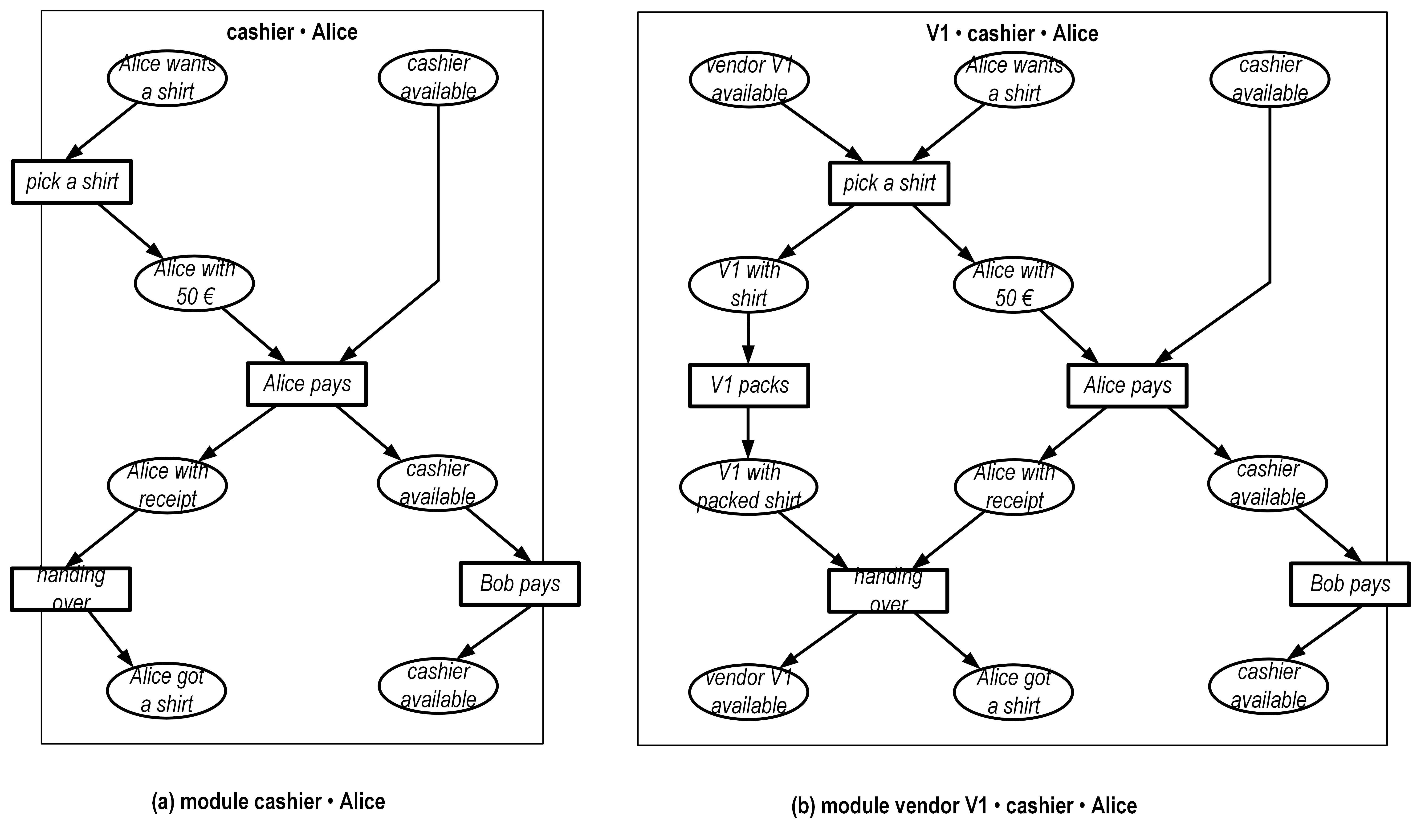}}
\caption{two further module compositions.}
\label{fig:composed_V1-cashier-Alice}
\end{figure}

We extend the example of Sec. 2.1 by modules for the remaining three agents of Fig.~\ref{fig:six_behaviours}, as in Fig.~\ref{fig:V1-cashier-Alice}. Fig.~\ref{fig:composed_V1-cashier-Alice} shows compositions of these modules. Interestingly, the module $\textit{cashier} \bullet \textit{Alice}$ in Fig.~\ref{fig:composed_V1-cashier-Alice}(a) is an example of a module with elements in both the left and right interfaces. Finally, the module in Fig.~\ref{fig:composed_V1-cashier-Alice}(b) composes all three modules.

\begin{figure}[!tb]
\centering
{\includegraphics[trim={0cm 0cm 0cm 0cm},clip,scale=.25]{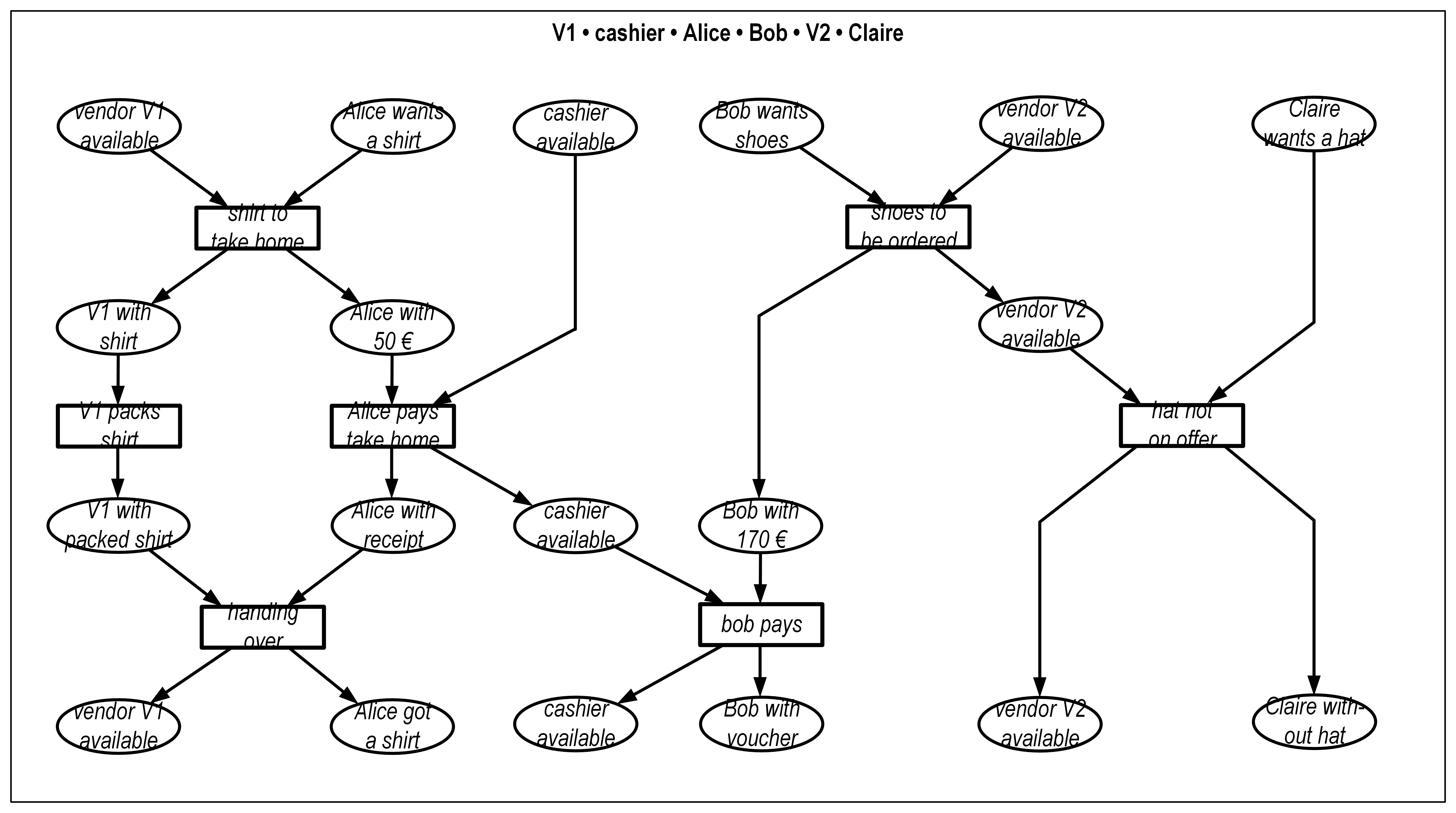}}
\caption{behavioral module $V \coloneqq V1 \bullet \textit{cashier} \bullet \textit{Alice} \bullet V2 \bullet \textit{Claire} \bullet \textit{Bob}$.}
\label{fig:overall_composition}
\end{figure}

We can now compose the composed module in Fig.~\ref{fig:composed_V1-cashier-Alice}(b) with the composed module in Fig.~\ref{fig:V2-Claire-Bob}(b), and obtain the composed module 
\begin{equation}
V \coloneqq V1 \bullet \textit{cashier} \bullet \textit{Alice} \bullet \textit{Bob} \bullet V2 \bullet \textit{Claire}
\end{equation}
in Fig.~\ref{fig:overall_composition}. The two interfaces of this module do not contain any elements.

From an abstract and more systematic point of view, the expression (3) is a bit unattractive. It would be nicer to have the module $trade \coloneqq V1 \bullet cashier \bullet V2$, with all interface elements on the right, and the module $customers \coloneqq Alice \bullet Claire \bullet Bob$ with all interface elements on the left. The module $V$ in (3) is then written as $trade \bullet customers$. Indeed, this is possible without any problems, because the modules \textit{Alice} and $V2$ have disjoint interfaces. According to equation (2), the sequence of the two modules $V2$ and \textit{Alice} in (3) can be swapped.

Summing up, the module $V$ of Fig.~\ref{fig:overall_composition} represents a typical single \textit{run} of a system. $V$ provides insight into subtle details of the mutual relationship of the events of the joint behavior of the involved six agents. For example, in the presented run, the joint events of the modules $V1$, \textit{Alice} and the \textit{cashier} are detached from the events of the modules of the other three agents. Bob waits until the cashier is finished with Alice. But vendor $V2$ and Alice are not related at all to the cashier. 

All this insight has been gained from the event log of Fig.~\ref{fig:event_log_neu}, together with the intuitively obvious idea that events of the business people will never be merged, hence they come with elements in right interfaces only, and correspondingly, events of the customers will never be merged, thus all come with elements in left interfaces. The choice of left and right interface is motivated by the dichotomy between shop modules (vendors and cashiers), and client modules. Of course, right and left may be swapped here. So, the interface elements of shop modules and of client modules complement each other.

\subsection{Composing an Occurrence Module From Occurrence Atoms}

Here we consider an alternative way of constructing occurrence modules.  In Sec.~2.3 we composed the run in Fig.~\ref{fig:overall_composition} from the modules of the six behavioral strands of agents, given in Fig.~\ref{fig:six_behaviours}.  Occurrence modules are frequently, but not always, composed from modules generated by such agents. Alternatively, an occurrence module can be generated from \textit{occurrence atoms}. An occurrence atom is a module that represents a single transition together with its surrounding arcs and places. We denote the occurrence atom of a transition $t$ by $\underline{t}$. To correspond to the previous representation of occurrence modules, we place the left interface of an occurrence atom at the top and the right interface at the bottom of its graphical representation.

\begin{figure}[!tb]
\centering
{\includegraphics[trim={0cm 0cm 0cm 0cm},clip,scale=.25]{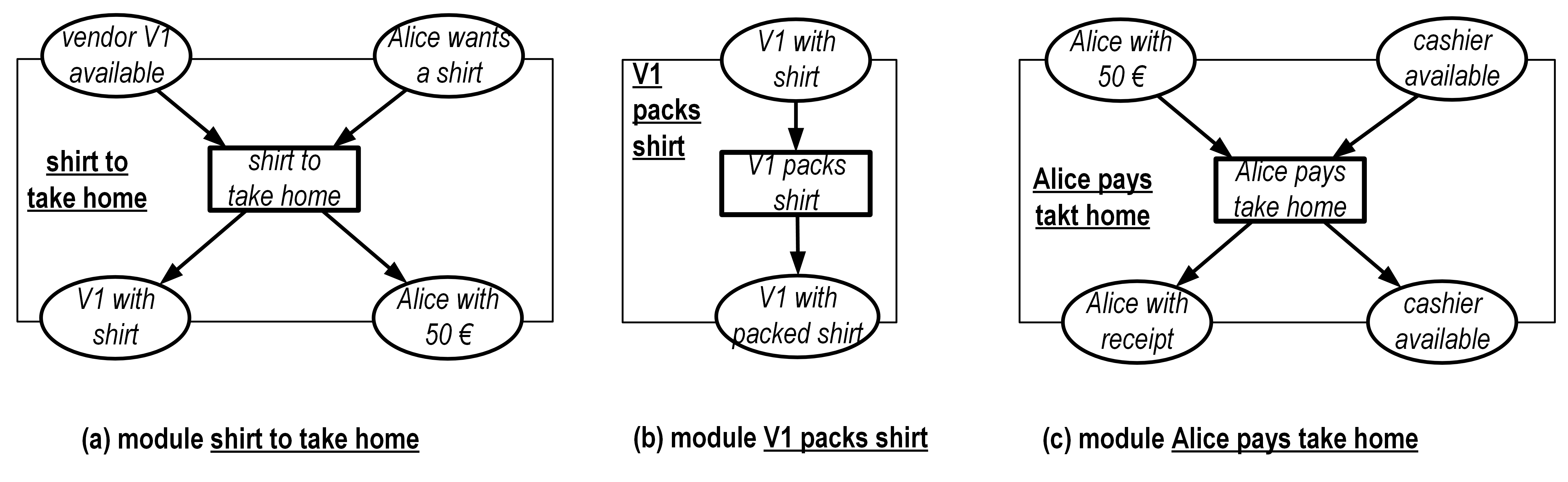}}
\caption{occurrence atoms}
\label{fig:occ_atoms}
\end{figure}

\begin{figure}[!tb]
\centering
{\includegraphics[trim={0cm 0cm 0cm 0cm},clip,scale=.25]{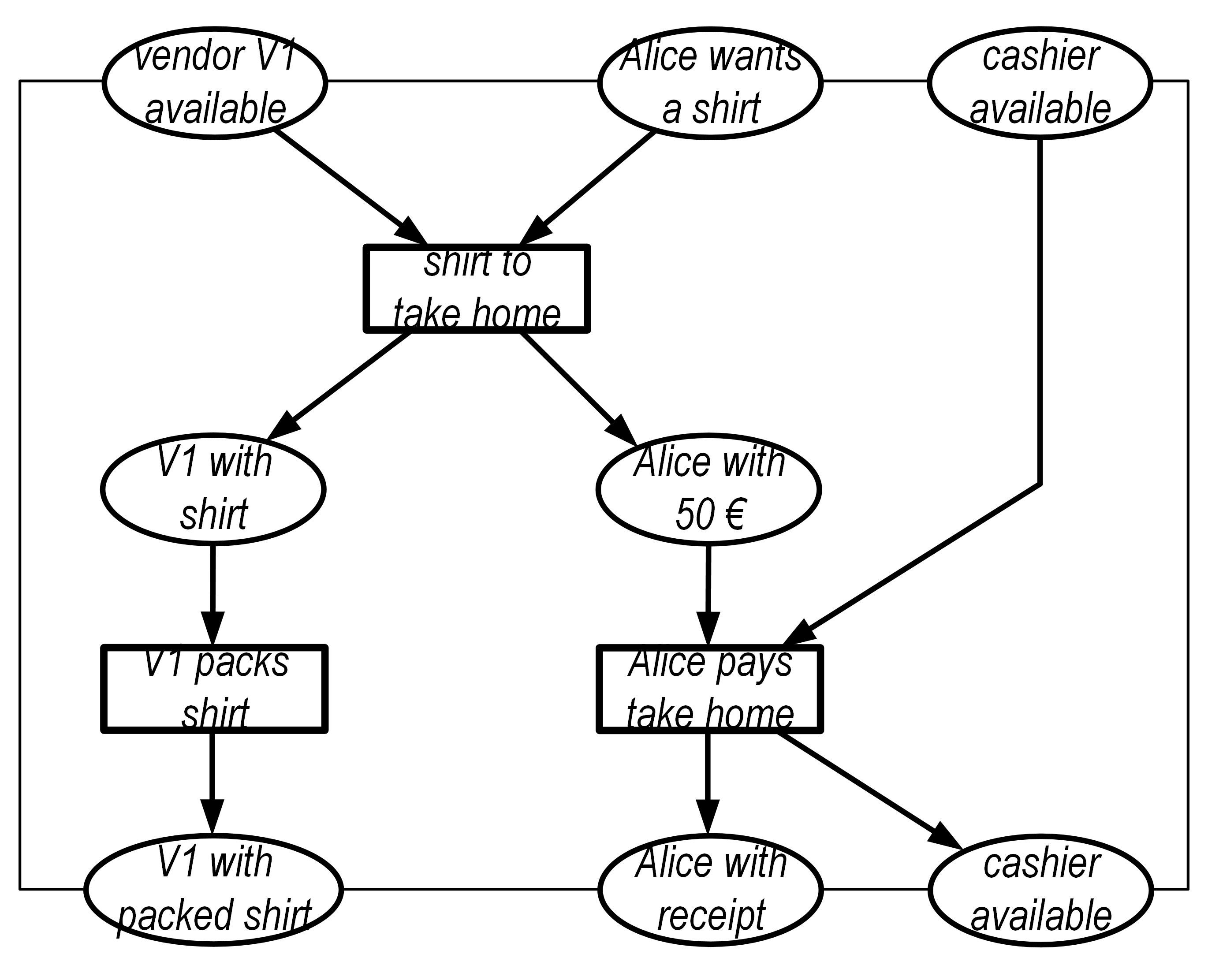}}
\caption{module M0: \textit{shirt to take home} $\bullet$ \textit{V1 packs shirt} $\bullet$ \textit{Alice pays take home}}
\label{fig:composition_of_atoms}
\end{figure}

Fig.~\ref{fig:occ_atoms}(a), (b) and (c) show the occurrence atoms of the transitions \textit{shirt to take home}, \textit{V1 packs shirt}, and \textit{Alice pays take home} of module $V$ of Fig.~\ref{fig:overall_composition}. The composition of the three occurrence atoms in Fig.~\ref{fig:composition_of_atoms} is identical to the upper left part of module $V$. It is easy to see how the occurrence atoms of the remaining four transitions of $V$ can be generated, and that their composition yields the entire module $V$. In fact, this is generally true: the occurrence atoms $\underline{t}$ of the transitions $t$ of an occurrence net $N$ can be arranged as a sequence $\underline{t_1}, \dots, \underline{t_n}$ such that 
\begin{equation}
N = \underline{t_1} \bullet \dots \bullet \underline{t_n}.
\end{equation}
This representation will be used in the following sections.

\section{System Models}
So far, we showed how to deduce a single run from a given event log. Our aim, however, is to deduce a system model from an event log. To this end, we need a manageable kind of system models. Here we derive such system models.

Many similar logs would yield many similar runs. Now we show how to extrude a system model from a set of runs. In a first step, we concentrate on the systematic management of involved data and functions.

\subsection{Structures and Signatures}
\begin{figure}[!tb]
\centering
\begin{minipage}{.5\textwidth}
{\includegraphics[trim={0cm 0cm 0cm 0cm},clip,scale=.25]{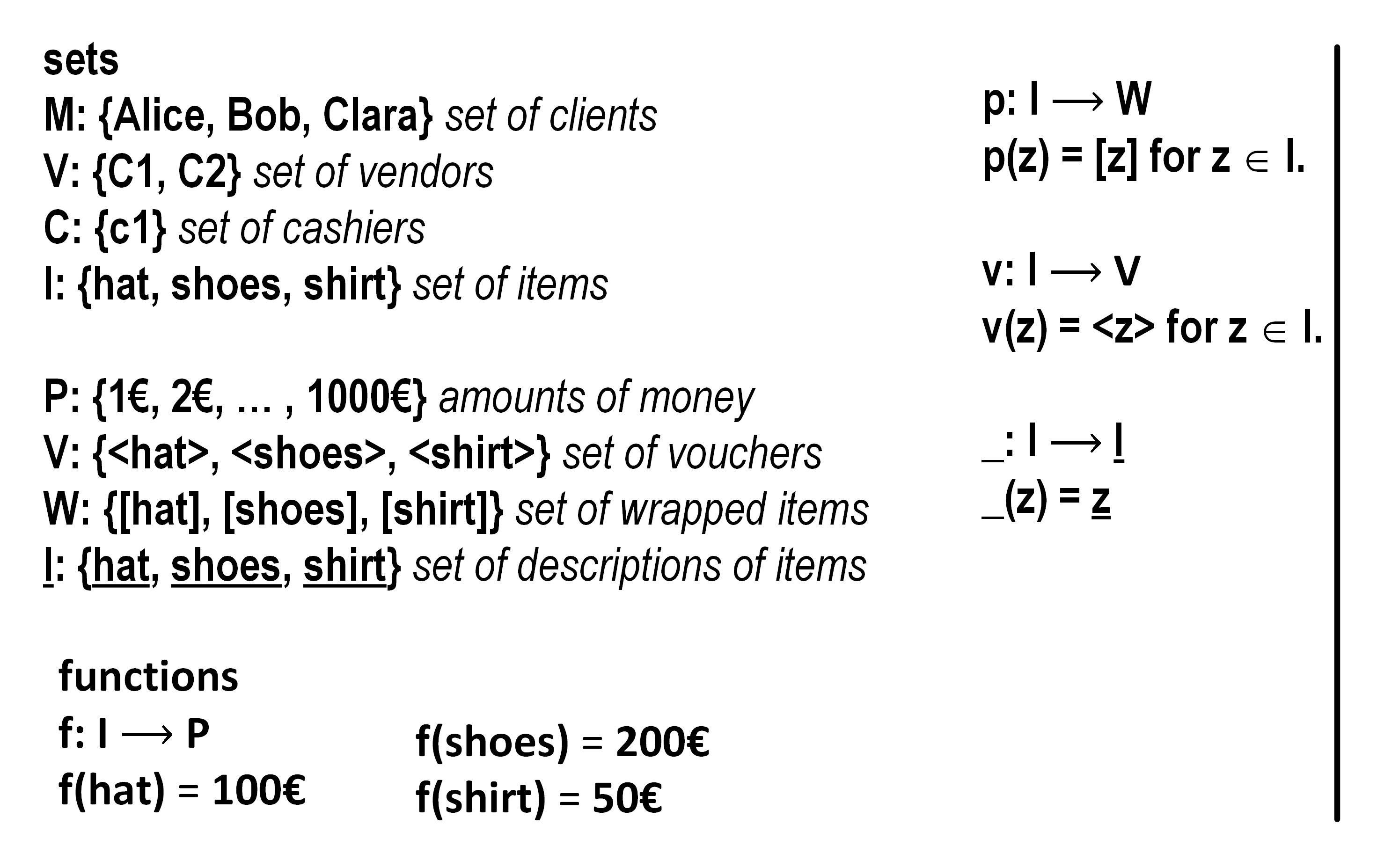}}
\caption{structure $S_0$, describing the system's data.}
\label{fig:structure_S0}
\end{minipage}%
\begin{minipage}{.5\textwidth}
{\includegraphics[trim={0cm 0cm 0cm 0cm},clip,scale=.25]{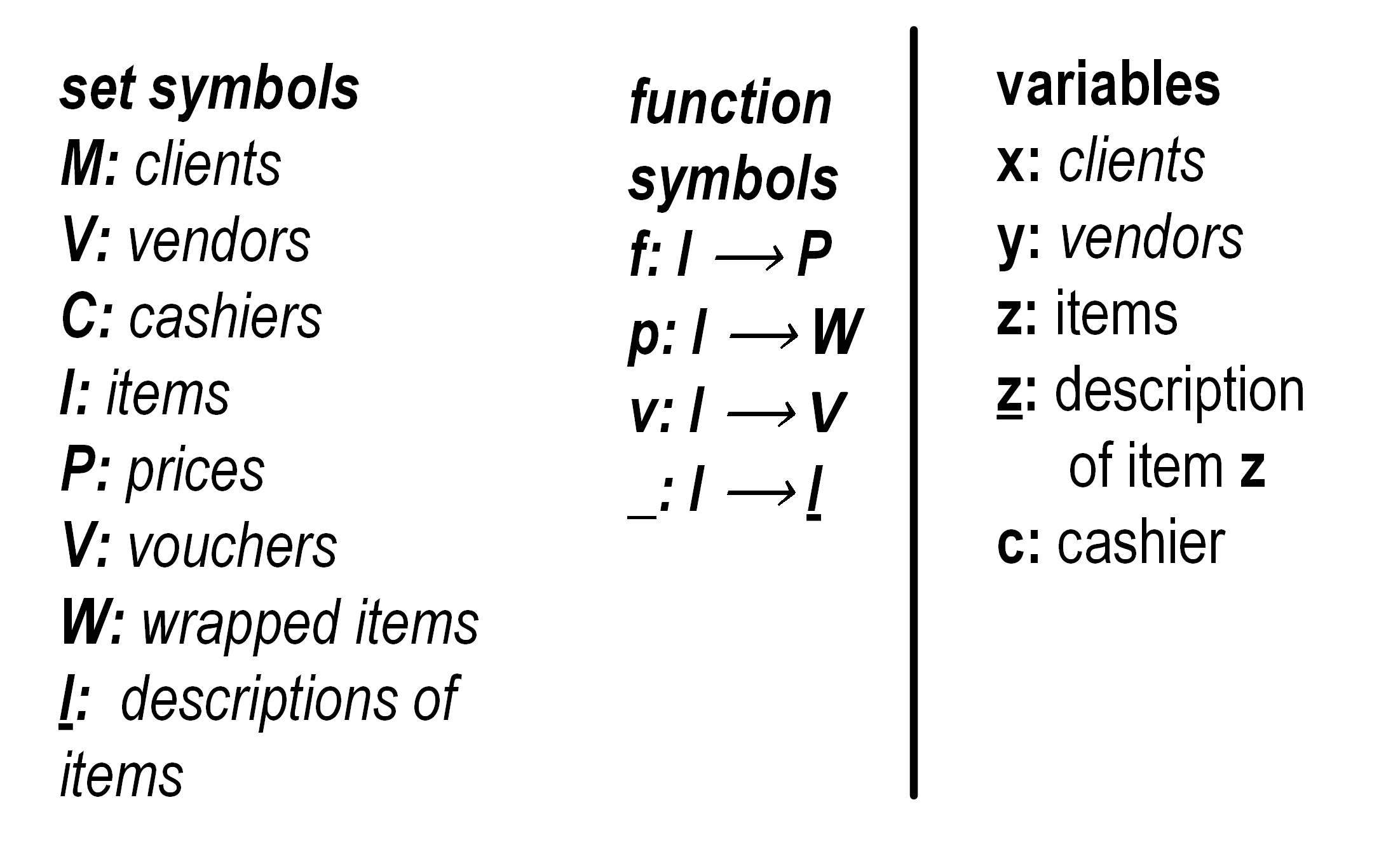}}
\caption{signature $\Sigma_0$ and variables for the structure $S_0$}
\label{fig:signature_Sigma}
\end{minipage}
\end{figure}

To cope with data and functions on data, we employ \textit{signatures} and \textit{signature-structures}, well-known in mathematics from general algebra and first order logic, and in informatics from algebraic specifications~\cite{sanella20212algebraic}. Fig.~\ref{fig:structure_S0} shows the signature-structure $S_0$ for the running example, consisting of eight sets and four functions. Each set is finite and includes real or imagined objects such as \textit{clients}, \textit{vendors}, \textit{cashiers}, \textit{products}, \textit{vouchers}, \textit{wrapped items}, but also more abstract items such as \textit{money} and \textit{descriptions of items}. In the course of systems mining, a structure like this should be provided by the provider of the logs. It may also be deducible from the logs. 

A symbolic representation of a system requires abstract, symbolic representations of structures such as in Fig.~\ref{fig:structure_S0}. This is achieved by means of signatures: a \textit{signature} $\Sigma_0$ for a structure $S$ includes \textit{sorted symbols}: a symbol for each set and each function of $S$. Fig.~\ref{fig:signature_Sigma} gives a signature, $\Sigma_0$, for the above structure $S_0$. For the sake of simple notation, for each set and each function of $S_0$ we write the corresponding symbol of $\Sigma_0$ in italic.

Additionally, Fig.~\ref{fig:signature_Sigma} shows sorted variables. Sorted symbols and variables yield \textit{terms}, such as $f(z)$, or tuples of terms, such as $(x, f(z))$.  A \textit{valuation} $\beta$ of the variables assigns to each variable $v$ an item $\beta(v)$ of the structure $S_0$. For example, with $\beta(x) = Alice$, $\beta(y) = V1$ and $\beta(z) = shirt$, the tuples $(y, z)$ and  $(x, f(z))$ yield in $S_0$ the tuples

\begin{equation}
\beta(y, z) = (V1, shirt) \text{ and } \beta(x, f(z)) = (Alice, 50 \text{\euro{}}).
\end{equation}

\subsection{System Atoms and Their Composition}
\begin{figure}[!tb]
\centering
{\includegraphics[trim={0cm 0cm 0cm 0cm},clip,scale=.25]{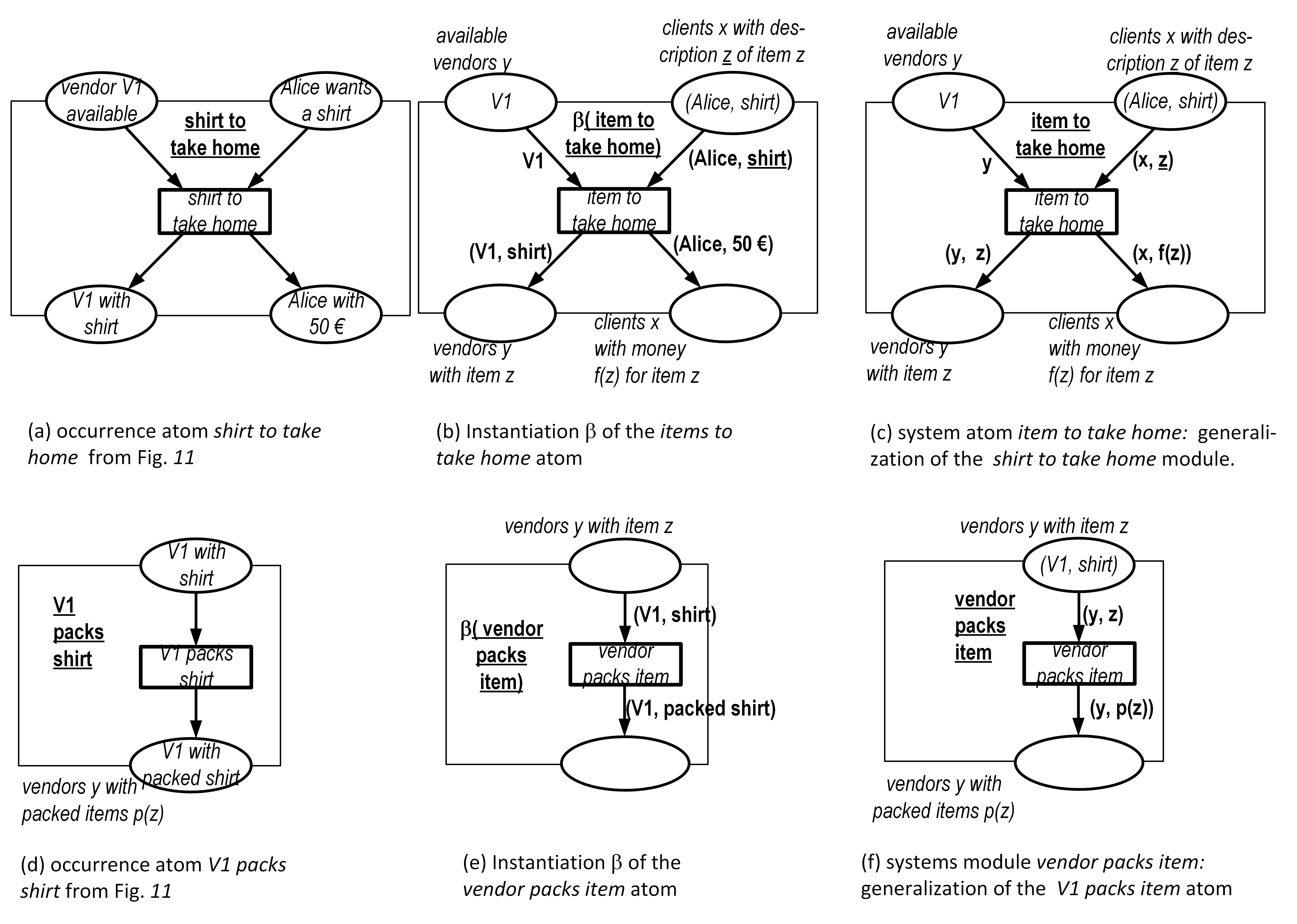}}
\caption{occurrence atoms and system atoms}
\label{fig:system_atoms}
\end{figure}

In order to extrude a system model from a set of runs, we start from single occurrence atoms, extruding a more general model of system atoms. Fig.~\ref{fig:system_atoms} shows an example: The atom \textit{\underline{shirt to take home}} of Fig.~\ref{fig:occ_atoms}(a) (repeated in Fig.~\ref{fig:system_atoms}(a)) is re-written in  Fig.~\ref{fig:system_atoms}(b): information about the vendor V1, the client Alice, the item shirt, and the price 50 \euro{} moves from the module’s places to its arcs. This representation is now conceived as an instantiation of the \textit{\underline{item to take home}} module in Fig.~\ref{fig:system_atoms}(c). In this module, the constant arc inscriptions of Fig.~\ref{fig:system_atoms}(b) are replaced by the variables $x$, $y$, and $z$, and terms $f(z)$ and $z$. In Figs.~\ref{fig:system_atoms}(b) and (c), the place inscriptions of the left (upper) interface places are conceived as tokens of the Petri net. Then, the firing rule of Petri nets defines the tokens for the right (lower) interface places. Fig.~\ref{fig:system_atoms}(b) is now gained as the instantiation of Fig.~\ref{fig:system_atoms}(c) by means of the above valuation $\beta$ as in (5). Of course, different valuations yield different instantiations of the \textit{\underline{item to take home}} system atom. This way, Fig.~\ref{fig:system_atoms}(c) is a system atom, representing many occurrence atoms.

In analogy to Fig.~\ref{fig:system_atoms}(a), (b), and (c), Fig.~\ref{fig:system_atoms}(d), (e), and (f), generalizes the \textit{\underline{V1 packs shirt}} occurrence atom as in Fig.~\ref{fig:occ_atoms}(b). It is obvious how from the five remaining occurrence atoms, the corresponding system atoms can be deduced. 

\begin{figure}[!tb]
\centering
{\includegraphics[trim={0cm 0cm 0cm 0cm},clip,scale=.25]{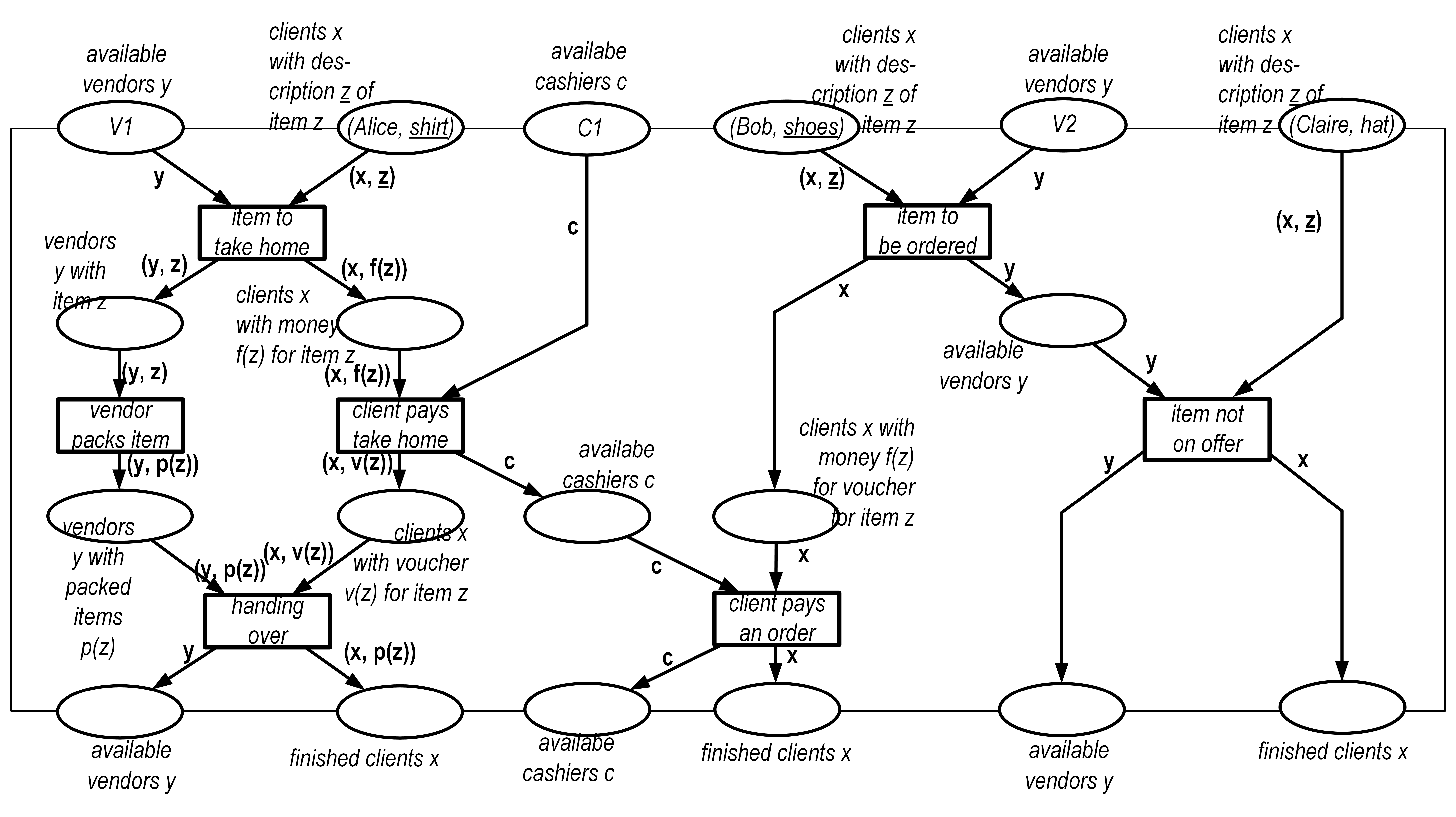}}
\caption{composed atoms: symbolic representation of the behavioral module $V$.}
\label{fig:composed_system_atoms}
\end{figure}

Fig.~\ref{fig:composed_system_atoms} composes the seven system atoms. This is a \textit{symbolic occurrence module}. Content wise, with the valuation $\beta$ for all arc inscriptions, it is just a different representation of the occurrence module $V$ in Fig.~\ref{fig:overall_composition}. Denotations of places and transitions have slightly been adjusted to better support intuition. The place inscriptions of $V$ are gained in Fig.~\ref{fig:composed_system_atoms} by the Petri net firing rule. Further, Fig.~\ref{fig:composed_system_atoms} has a non-empty left and right interface, in contrast to Fig.~\ref{fig:overall_composition}.

\subsection{Constructing a System Net From Symbolic Occurrence Modules}
\begin{figure}[!tb]
\begin{minipage}{.6\textwidth}
\centering
{\includegraphics[trim={0cm 0cm 0cm 0cm},clip,scale=.2]{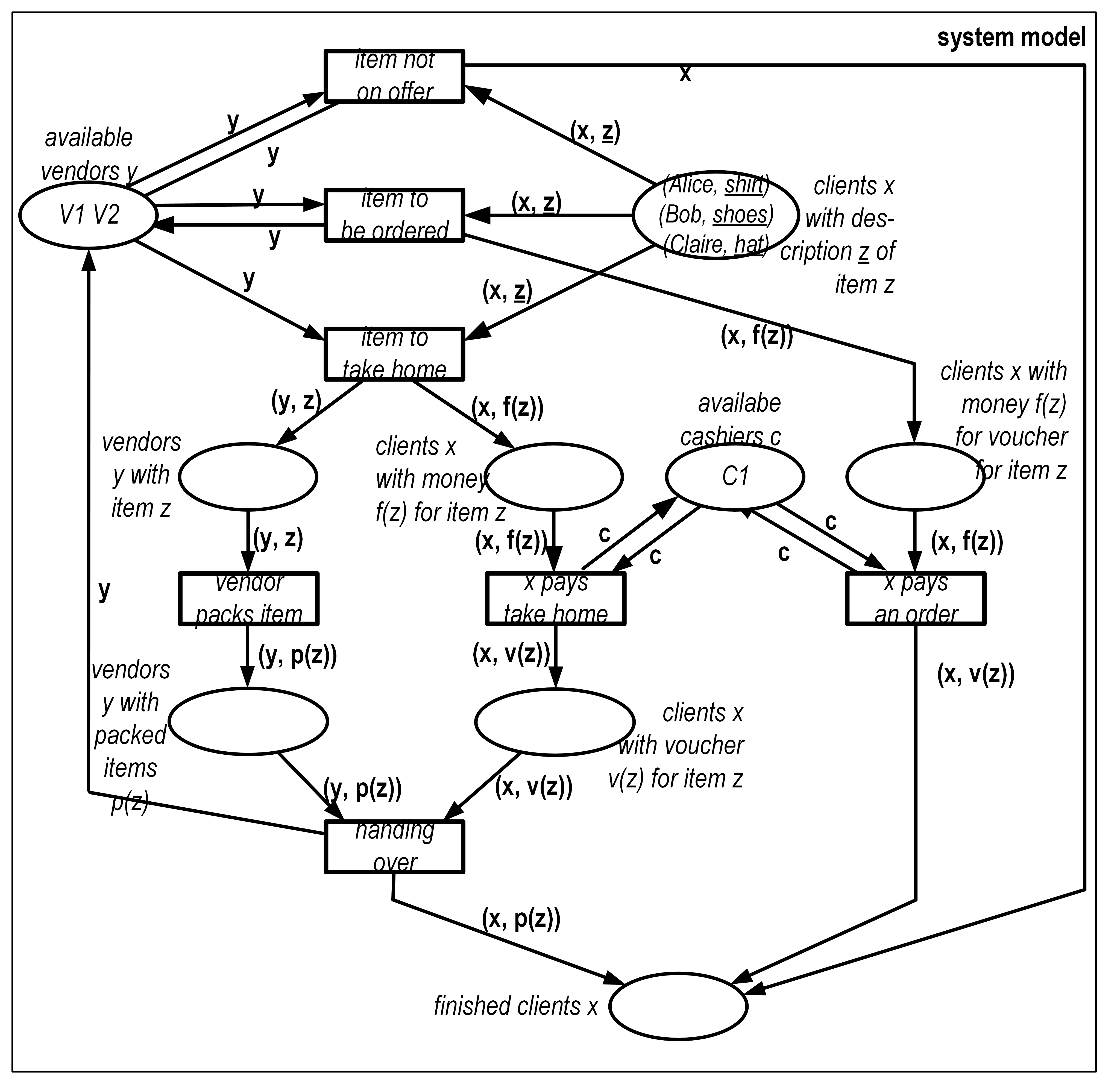}}
\caption{system model $M$.}
\label{fig:system_model}
\end{minipage}
\begin{minipage}{.6\textwidth}
{\includegraphics[trim={0cm 0cm 0cm 0cm},clip,scale=.2]{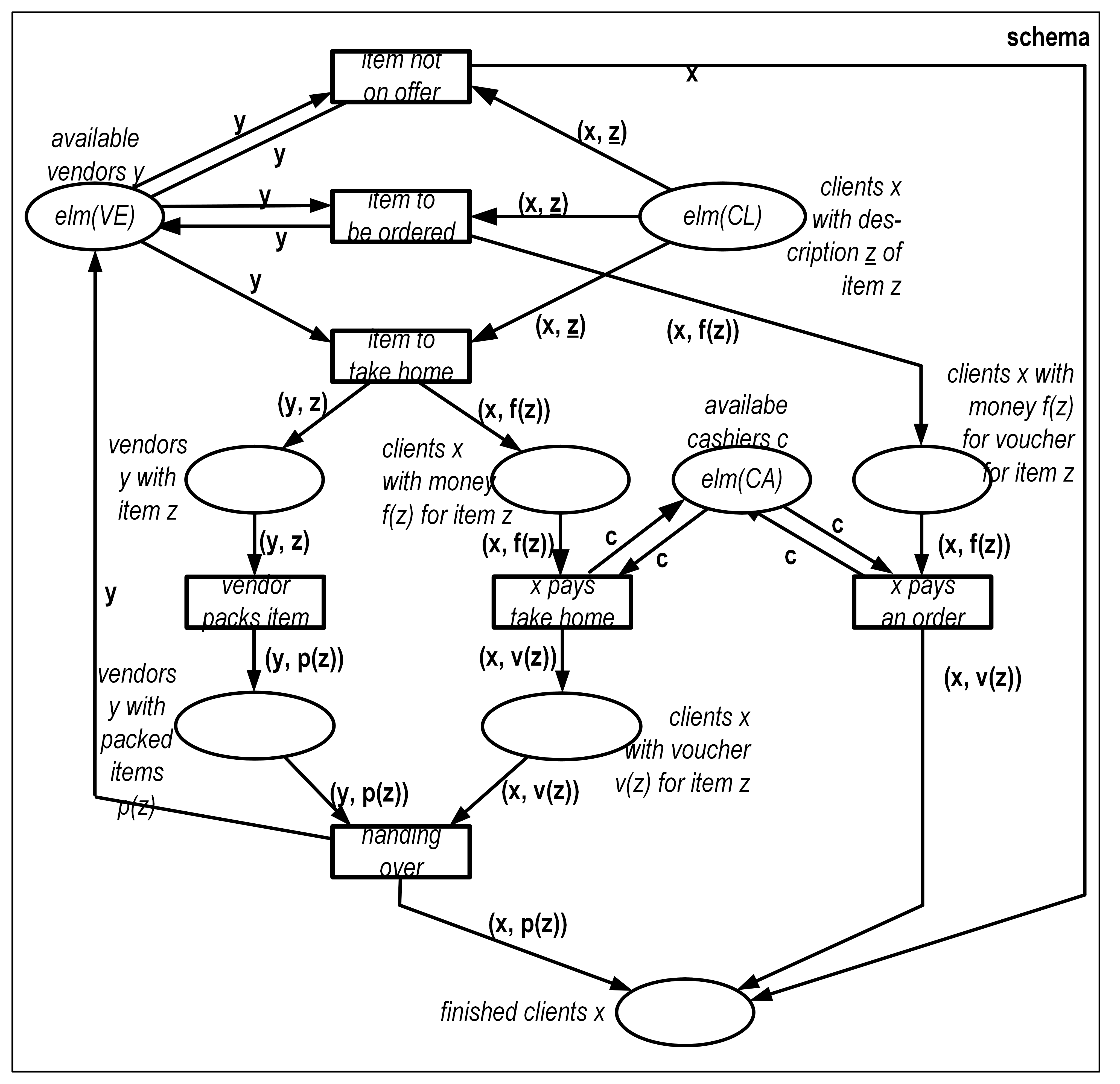}}
\caption{schema: symbolic initial marking.}
\label{fig:schema}
\end{minipage}
\end{figure}

It is now possible to deduce a full-fledged Petri net model from the symbolic occurrence module of Fig.~\ref{fig:composed_system_atoms}: Just identify equally labeled places. The resulting system net is shown in Fig.~\ref{fig:system_model}. The tokens of Fig.~\ref{fig:composed_system_atoms} are collected on the corresponding place of Fig.~\ref{fig:system_model}. 

Fig.~\ref{fig:system_model} shows a high level Petri net. It specifies a lot of runs, depending on the choice of the valuation $\beta$ of the variables. Furthermore, now, even when fixing the valuation $\beta$ as above, each client with his description of an item may now execute any of the three events \textit{item not on offer}, \textit{item to be ordered} and \textit{item to take home}, with any of the two vendors $V1$ or $V2$. This is a generalization that suggests itself from the assumptions of the system.

\subsection{Deriving a Net Schema}
The system model in Fig.~\ref{fig:system_model} fixes the sets of vendors, clients, and items. One would prefer a specification that leaves these sets open, to be fixed as an \textit{interpretation} of those symbols by the user of the model. For this purpose, it suggests itself to use fresh symbols, e.g. ${V \! E}$, ${C \! L}$, and ${C \! A}$, to be interpreted as sets of vendors, clients, and cashiers, as initial tokens on the places \textit{available vendors}, \textit{clients with descriptions of items}, and \textit{available cashiers}, resp. However, this is not exactly what we want: An interpretation of ${V \! E}$ would, for example, interpret the symbol ${V \! E}$ by the set $\{V1, V2\}$ as one token on the place \textit{available vendors}. Instead, we want two tokens, $V1$ and $V2$. This is represented by means of the “elm”-notation, as in the net of Fig.~\ref{fig:schema} (more details in \cite{fettke2021handbook}).

\section{How to Mine a System Model}
The notions, concepts, and constructs described in the above sections suggest to mine a system model starting from information on static system aspects such as the architecture of the system, the data structures, and the involved agents. The data and the operations on the data are systematically represented in a signature-structure. The architecture and the agents provide the background for modeling dynamic aspects, i.e. for the derivation of occurrence modules as models for runs, and finally system modules as models for the entire system.

\subsection{From Logs to Runs}
The first step identifies for each agent its sequential behavior from the log, and constructs a distributed run from the agents’ behaviors:  

\begin{enumerate}

\item From a given event log, for each agent identify in the log the events which involve the agent. The sequence of those events constitute the \textit{behavior} of the agent in the log. Figs.~\ref{fig:event_log_neu} and \ref{fig:six_behaviours} show corresponding examples. 

\item Turn the behavior of each agent into an occurrence module: each event either belongs to the module’s interior part, or its left or its right interface. For an element, adequate choice of the interface depends on the intended composition with elements of other modules. Fig.~\ref{fig:V2-Claire}(a), and (b), ~\ref{fig:V2-Claire-Bob}(a), and \ref{fig:V1-cashier-Alice}(a), (b), and (c) show examples. 

\item Compose the agents’ occurrence modules: In general, an event of an event log is involved in more than one agents’ behavioral module. Composition of the modules yields a comprehensive occurrence module, i.e.  partially ordered run, as in Fig.~\ref{fig:overall_composition}.   

\end{enumerate}

\subsection{From Runs to Systems}

The second step identifies for each occurrence atom of a given partially ordered run a system atom with terms over the given signature-structure as arc inscriptions. From this representation, the sought system model is derived: 

\begin{enumerate}

\setcounter{enumi}{3}

\item For each occurrence atom of the run, identify the involved agents and data structures. Move this information from place inscriptions to arc inscriptions. The arc inscriptions then are terms of the underlying signature structure. Fig.~\ref{fig:system_atoms} shows examples. 

\item In this representation of each occurrence atom, replace each constant symbol by a variable. This yields a system atom.    

\item Compose those system atoms, as in Fig.~\ref{fig:composed_system_atoms}. 

\item In this representation, merge equally denoted places. This yields the sought system model, as in Fig.~\ref{fig:system_model}. 

\item To achieve a purely schematic representation, replace the initial marking by a symbolic marking, as in Fig.~\ref{fig:schema}.   

\end{enumerate}


\section{Related Work}

The main concepts for the theoretical foundations of process mining are based on the idea of grammar inference, grammar induction, or language identification \cite{higuera2010grammatical_inference}, which was originally proposed by \cite{gold1967}. Since these theoretical models do not adequately represent all interesting aspects of business processes, a plethora of enhanced formal frameworks are developed \cite{aalst2016mining}. However, none of these approaches are completely satisfactory because the role of causality, subsystems, and data are not integrated and adequately covered. Besides the theoretical work, many practical approaches originate from engineering process modeling and mining systems \cite{gartner2019market_guide}. However, these approaches lack a theoretical foundation.

Recent work in the area of artifact-centric \cite{Fahland2019}, object-centric process mining \cite{aalst2020object_centric}, and agent system mining \cite{tour2021agent}, addresses these lacks already. Although these ideas clearly show improvements compared to the classical understanding of systems, models, and logs as formal languages, they still do not provide a satisfactory understanding of system architecture and the difference of abstract and concrete data structures which are strongly needed for an integrated understanding of business systems. Additionally, our understanding of an agent is rather general compared to the technical notion used by \cite{tour2021agent}.

Although recent work acknowledges the need for representing causal structures, the choice is often not satisfactory. C.A. Petri formulated the concept of distributed runs as early as the late 1970s \cite{petri1977non_sequential}. It has been taken up again and again, also under the names "true concurrency", or "partial order semantics", but initially did not prevail over sequential processes. One of the reasons for this was the comparatively complex technical apparatus for dealing with distributed processes, combined with comparatively little benefit. Meanwhile, the basic ideas of distributed runs are used in many contexts, e.g. partial order process mining \cite{vanderaa2020partial_order}. Furthermore, the composition calculus, as used in this contribution, provides adequate and simple technical tools.

To cope properly with data aspects, and in particular to properly integrate behavioral and data aspects in one formal framework, we resort to signature-structures, the established formal basis of first order logic and algebraic specifications \cite{sanella20212algebraic}. 
Models of really big systems are gained by composing models of subsystems. The composition calculus covers also this aspect, as developed in \cite{fettke2021handbook}.

\section{Conclusion}

Classical process mining assumes a run as a sequence of events and then tries to solicit information about concurrent, independent event occurrences $a$ and $b$ form the observation, that in many similar logs, $a$ and $b$ occur in either order. We suggest to start considering a run as an unordered set of events, and then to order them, as much as reasonable, by considering agents and the composition of agents’ behavior. For example, the module $V$ of Fig.~\ref{fig:overall_composition} provides insight into subtle details of the mutual relationship of the events of the joint behavior of the involved six agents. In the presented run, the joint events of the modules \textit{vendor V1}, \textit{Alice} and \textit{cashier} are detached from the events of the modules of the other three agents. Bob waits until the cashier is finished with Alice. But vendor $V2$ and Alice are not related at all to the cashier. All this has been gained from the event log of Fig.~\ref{fig:event_log_neu}, together with the intuitively obvious idea that the events of the trade components never should be merged, hence all go to the right interfaces, and correspondingly the events of the customers should never be merged, thus all go to the left interfaces. Of course, right and left may be swapped here.

In this paper, we argue that \textit{causality, composition}, and \textit{objects matter} while mining a system. We introduce the foundational concepts for conducting system mining. In the future, more case studies need to be done and new tools for supporting the main ideas of \textsc{Heraklit} have to be developed. So, in the future, we speculate that the two academic worlds of data and process mining will be complemented with and enhanced by systems mining allowing a deeply integrated understanding of business processes.

%
%
%

\bibliographystyle{splncs04}
\bibliography{main}





\end{document}